# EEG-assisted retrospective motion correction for fMRI: E-REMCOR

Vadim Zotev*, Han Yuan, Raquel Phillips, Jerzy Bodurka*

Laureate Institute for Brain Research, 6655 S. Yale Avenue, Tulsa, OK 74136, USA

**Abstract**

We propose a method for retrospective motion correction of fMRI data in simultaneous EEG-fMRI that employs the EEG array as a sensitive motion detector. EEG motion artifacts are used to generate motion regressors describing rotational head movements with millisecond temporal resolution. These regressors are utilized for slice-specific motion correction of unprocessed fMRI data. Performance of the method is demonstrated by correction of fMRI data from five patients with major depressive disorder, who exhibited head movements by 1−3 mm during a resting EEG-fMRI run. The fMRI datasets, corrected using eight to ten EEG-based motion regressors, show significant improvements in temporal SNR (TSNR) of fMRI time series, particularly in the frontal brain regions and near the surface of the brain. The TSNR improvements are as high as 50% for large brain areas in single-subject analysis and as high as 25% when the results are averaged across the subjects. Simultaneous application of the EEG-based motion correction and physiological noise correction by means of RETROICOR leads to average TSNR enhancements as high as 35% for extended brain regions. These TSNR improvements are largely preserved after the subsequent fMRI volume registration and regression of fMRI motion parameters. The proposed EEG-assisted method of retrospective fMRI motion correction (referred to as E-REMCOR) can be applied to improve quality of fMRI data with severe motion artifacts and to reduce spurious correlations between the EEG and fMRI data caused by head movements. It does not require any specialized equipment beyond the standard EEG-fMRI instrumentation and can be applied retrospectively to any existing EEG-fMRI data set.

## 1. Introduction

Simultaneous EEG-fMRI, which combines the advantages of high spatial resolution of fMRI and high temporal resolution of EEG, has evolved into a powerful and widely used neuroimaging method (Mulert and Lemieux, 2010). However, both fMRI and concurrent EEG suffer from physiological confounds, particularly head motions. Children, elderly people, patients with mental disorders (or other medical conditions), and even healthy controls engaged in demanding experimental tasks – can all exhibit significant head movements. The resulting motion artifacts, if not properly corrected, can severely reduce the quality of both fMRI and EEG data and even make the fMRI and EEG data sets unusable. Therefore, implementation of efficient motion correction techniques is particularly important for simultaneous EEG-fMRI.

In addition to the overall reduction in fMRI and EEG data quality, head motions can introduce systematic effects that may influence interpretation of the neuroimaging data on the group level. In task fMRI, head movements often correlate with experimental tasks, thus affecting the ability of fMRI to detect task-related neuronal activity (e.g. Hajnal et al., 1994; Johnstone et al., 2006). In fMRI studies of resting-state functional connectivity, even small (<0.5 mm) head motions have been shown to cause spurious correlations in group-level functional connectivity networks (Power et al., 2012; Van Dijk et al., 2012). In EEG-fMRI, it has been shown that motion-related EEG artifacts, when convolved with the hemodynamic response function and correlated with fMRI time courses, predict "neuronally plausible" patterns of activation in the motor areas (Jansen et al., 2012). Careful examination and correction of such motion-induced correlations between EEG and fMRI data are particularly important for EEG-fMRI studies of spatiotemporal brain dynamics (e.g. Britz et al., 2010; Yuan et al., 2012).

The basic motion correction step in fMRI data processing is volume registration, which aligns each 3D multislice brain image (referred to as volume) in the fMRI time series with the reference brain image by means of a rigid-body spatial transformation with six motion parameters (e.g. Friston et al., 1995; Jiang et al., 1995; Cox and Jesmanowicz, 1999). The volume registration alone cannot remove all motion artifacts in fMRI data for the following reasons: i) head motions can cause spatially varying spin warping by changing spatial distributions of local magnetic susceptibility gradients (Jiang et al., 1995); ii) head motions can change spin excitation history, especially if they involve movements of tissue through the fMRI slice plane (Friston et al., 1996); iii) head motions, occurring on time scales shorter than the fMRI repetition time $TR$, can affect only some of the slices within a volume. These effects, together with fMRI signal variations due to cardiac and respiratory activity, make head movements appear non-rigid in fMRI images. They can severely reduce the performance of the volume registration in addition to possible registration inaccuracies due to interpolation effects (Grootoonk et al., 2000). The next common motion correction step is application of a linear regression procedure with the six motion parameters as regressors to reduce residual motion-related signal variations in fMRI time courses (e.g. Friston et al., 1996; Johnstone et al., 2006).

---

* Corresponding authors.
*E-mail address*: vzotev@laureateinstitute.org (V. Zotev);
jbodurka@laureateinstitute.org (J. Bodurka).

Head motions impair the ability of BOLD fMRI to detect BOLD signal variations due to neuronal activity by reducing temporal SNR (TSNR) of fMRI time series (Murphy et al., 2006). TSNR is defined as a ratio of mean fMRI signal to its temporal standard deviation (see Methods below). Because physiological BOLD signal fluctuations increase in proportion to image SNR (Krüger and Glover, 2001), TSNR asymptotically reaches a maximum as the image SNR is improved. While the image SNR levels for echo-planar imaging (EPI) sequences on modern 3 tesla MRI scanners with receive head coil arrays can exceed 300, the maximum achievable TSNR values for gray matter are around 80-90 (Krüger and Glover, 2001; Bodurka et al., 2007) in the absence of head motions. These TSNR values can be further substantially reduced by random head movements. Therefore, improvement in TSNR of fMRI time series through efficient correction of motion effects is essential for maximizing fMRI sensitivity to neuronal activity and reducing fMRI scan duration.

Another complication caused by random head movements in fMRI is that the effects of such movements are superimposed on the effects of physiological motions due to cardiac and respiratory processes. Large head movements can prevent accurate removal of cardiorespiratory artifacts by means of physiological noise correction methods such as RETROICOR (Glover et al., 2000). This method approximates the cardiorespiratory effects in fMRI time series by Fourier regressors depending on cardiac and respiratory phases. It has been suggested (Jones et al., 2008) that, in order to improve accuracy of the cardiorespiratory correction in the presence of random head motions, volume registration should be performed before the application of RETROICOR. The proposed motion-corrected version of RETROICOR (Jones et al., 2008), however, requires an individual set of regressors with different time courses for each voxel, and can take into account only those random motion effects that are correctable by volume registration.

It has been long recognized that accurate motion correction of fMRI data, particularly in the widely used EPI imaging, requires measurements of head movements with high temporal resolution. Various approaches have been implemented to measure six rigid-body head motion parameters with fast temporal sampling independently of the volume registration. They include the use of navigator echoes (e.g. Fu et al., 1995; Welch et al., 2002), optical tracking devices (e.g. Tremblay et al., 2005; Zaitsev et al., 2006, Qin et al., 2009), or RF microcoils as active markers (e.g. Dumoulin et al., 1993; Ooi et al., 2011). These techniques can be used for both retrospective and real-time prospective motion correction (Ward et al., 2000). However, they all require a complex setup (e.g. specialized motion-sensitive equipment, hardware modifications to the MRI scanner) and/or specialized pulse sequences. They also have their shortcomings: the techniques based on navigator echoes or active markers increase the image acquisition time, while the optical tracking approaches require unobstructed lines of sight between the head and the tracking devices.

As with fMRI, head movements pose a major problem for EEG performed simultaneously with fMRI. Motion artifacts in EEG data, recorded inside an MRI scanner, appear because of the existence of conductive loops along the surface of the head due to electrical conductivity of the scalp between EEG electrodes. Rotational movements of such loops in the static main magnetic field of the scanner induce time dependent artifact voltages picked up by EEG electrodes (see Methods below). These artifact voltages can exceed EEG signals due to neuronal activity and overlap with essential frequency bands of EEG spectrum, including the alpha band (8−13 Hz). Cardioballistic artifacts, caused by rapid head movements following the cardiac pulses, are one common example of motion artifacts in EEG-fMRI. Because they are quasi-periodic in time, the cardioballistic artifacts can be removed from the EEG data using techniques such as the average artifact subtraction method (Allen et al., 1998).

EEG artifacts due to random head movements are more difficult to characterize and correct. Approaches have been proposed for real-time adaptive filtering of both cardioballistic and random-motion artifacts using external motion monitoring by means of either a piezoelectric motion sensor (Bonmassar et al., 2002) or a special head cap with several wire loops (Masterton et al., 2007). Independent component analysis (ICA) (e.g. Bell and Sejnowski, 1995; Hyvärinen and Oja, 2000) is a powerful statistical method for separating signals from different sources in multichannel EEG recordings (e.g. Makeig et al., 1997). It has been successfully used to identify and remove various types of artifacts from EEG data in offline data analysis (e.g. Nakamura et al., 2006; Mantini et al., 2007).

Until the present, motion correction efforts for fMRI and for EEG concurrent with fMRI have followed two separate, though somewhat parallel, paths. No attempt has been made to combine the motion correction approaches developed separately for fMRI and EEG. Yet, the EEG array is itself a sensitive detector of head motions inside an MRI scanner, capable of tracking rapid head movements with millisecond temporal resolution.

Here we propose a novel EEG-based method for retrospective fMRI motion correction that uses motion artifacts in EEG data, recorded simultaneously with fMRI, to generate high-temporal-resolution motion regressors for fMRI. We refer to it as E-REMCOR (EEG-assisted REtrospective Motion CORrection). We demonstrate the efficiency of E-REMCOR in five patients with major depressive disorder (MDD), who exhibited head movements by 1−3 mm during a nine-minute-long resting EEG-fMRI run.

## 2. Methods

*2.1 E-REMCOR*

The main reason for the appearance of artifacts in EEG data recorded inside an MRI scanner is the presence of an electrical (ionic) conductivity path between any pair of EEG electrodes. Because conductivity of the scalp tissue is about 20 times higher than that of the skull, it is sufficient to consider conductivity paths within the scalp, i.e. along the surface of the head. This is a simple model that captures the essential physics of the problem (Nakamura et al., 2006). An accurate electromagnetic analysis of EEG artifacts would require modeling several tissue layers with different electrical conductivities, including scalp, skull, CSF, and brain (e.g. Nunez and Srinivasan, 2006, chapters 4,6).

A change in magnetic flux penetrating a contour, formed by the conductive path together with EEG electrodes' leads, induces an electromotive force (EMF) in the contour, as described by Faraday's law. Because EEG signals are typically measured with respect to a single reference electrode (Ref), an effective contour for each EEG channel includes both that channel's electrode and Ref, but the contour's precise size and shape are unknown. If the EEG amplifier inputs draw no current, the voltage measured by a given EEG channel, $V_{EMF}$, is equal to the EMF itself. Rapidly changing magnetic fields due to RF pulses and switching gradients applied during an fMRI sequence produce MRI artifacts in the EEG data. Head motions in the scanner's main magnetic field, caused by cardiac pulsations, induce cardioballistic artifacts. Both types of artifacts are reasonably well understood and can be efficiently (though not completely) removed from EEG data acquired simultaneously with fMRI (e.g. Allen et al., 1998, 2000).

In this work, we focus on EEG artifacts resulting from random head movements in the static uniform magnetic field of the MRI scanner. $V_{EMF}(t)$ refers to the voltage of such artifact measured by a given EEG channel. According to Faraday's law, $V_{EMF}(t) = -d\Phi/dt$, so one can write:

$$\int_0^t V_{EMF}(\tau)d\tau = \Phi(0) - \Phi(t) = \Delta\Phi_R(t) + \Delta\Phi_D(t) \quad (1)$$

Here $\Phi(t)$ is magnetic flux penetrating a given EEG channel's effective contour at time $t$, and $\Phi(0)$ is the flux at $t=0$. The term $\Delta\Phi_R(t)$ describes the flux change due to rotations of the contour without deformations, while $\Delta\Phi_D(t)$ accounts for possible small contour deformations in addition to rotations. Because the MRI scanner's main magnetic field (magnetic flux density) $B_0$ is highly uniform, a parallel translation of a constant-geometry contour in any direction will not induce any EMF. For an arbitrary-shape contour of surface area $A$, the rotational part of the flux change can be expressed by a surface integral as follows:

$$\Delta\Phi_R(t) = B_0 \int_A \mathbf{n}_B \cdot [\mathbf{n}(s,0) - \mathbf{n}(s,t)] da_s \quad (2)$$

Here $da_s$ is an elementary flat area (labeled by a variable $s$) on the surface spanning the contour, $\mathbf{n}(s,t)$ is a unit normal vector to this area at time $t$, $\mathbf{n}(s,0)$ is the normal vector at $t=0$, $\mathbf{n}_B$ is a unit vector along $B_0$, and "·" is a scalar product of two vectors.

In functional MRI, small rigid-body motions of the head are commonly described by six motion parameters defined in the image reference frame (Jiang et al., 1995). The first three parameters are translations along x (left-right), y (posterior-anterior), and z (inferior-superior) axes of the reference brain image. The other three parameters are rotations around x axis ($\varphi$, pitch), y axis ($\psi$, roll), and z axis ($\theta$, yaw). A general head rotation is described by the following 3D rotation matrix (Jiang et al., 1995), which is a product of three matrices, $\mathbf{R}_x(\varphi)$, $\mathbf{R}_y(\psi)$, and $\mathbf{R}_z(\theta)$, corresponding to the three basic rotations:

$$\mathbf{R}(\varphi,\psi,\theta) = \mathbf{R}_z(\theta(t)) \cdot \mathbf{R}_x(\varphi(t)) \cdot \mathbf{R}_y(\psi(t)) \quad (3)$$

With the assumptions that the effective contour rotates rigidly with the head and the reference brain image corresponds to $t=0$, Eq (2) can be re-written using the rotation matrix $\mathbf{R}$ and the identity matrix $\mathbf{I}$:

$$\Delta\Phi_R(t) = B_0 \mathbf{n}_B \cdot [\mathbf{I} - \mathbf{R}(\varphi,\psi,\theta)] \int_A \mathbf{n}(s,0) da_s \quad (4)$$

The last expression shows that the temporal change in magnetic flux, $\Delta\Phi_R(t)$, which can be determined from the EEG motion artifact according to Eq (1), is, in general, a function of all three fMRI rotational motion parameters. If $\mathbf{n}_B$ is collinear with the image z axis, the flux change will only depend on $\varphi$ and $\psi$. This is a consequence of the fact that rotation of any vector around $B_0$ (in this case by angle $\theta$, yaw) does not change its projection on $B_0$. For small rotation angles, a power series expansion of $\mathbf{R}(\varphi,\psi,\theta)$ to the lowest (first) order in $\varphi$, $\psi$, and $\theta$ yields the following linear approximation with constants $c_1$, $c_2$, and $c_3$ depending on $A$, $B_0$, and the vectors $\mathbf{n}(s,0)$ and $\mathbf{n}_B$ in the image reference frame:

$$\Delta\Phi_R(t) \approx c_1\varphi(t) + c_2\psi(t) + c_3\theta(t), \quad \varphi,\psi,\theta \ll 1 \qquad (5)$$

Equations (4) and (5) suggest that, if geometrical properties of three different contours and their orientations at $t=0$ are known precisely, the rotational motion parameters $\varphi(t)$, $\psi(t)$, and $\theta(t)$ can be accurately determined for any $t$ using simultaneous measurements of $\Delta\Phi_R(t)$ functions for those three contours. For motion artifacts, recorded by different EEG channels, the effective contour properties are unknown. Nevertheless, if motion artifacts dominate the EEG recordings, $\Delta\Phi_R(t)$ functions, obtained according to Eq (1), can be used as motion regressors, because they represent different linear combinations of the actual motion parameters. Selection of EEG channels for generation of such regressors is rather ambiguous, however.

For the purpose of fMRI motion correction (based on the EEG data after the removal of MRI and cardioballistic artifacts), it is important to separate EEG artifacts arising due to random head motions, $V_{EMF}(t)$, from other instrumental and physiological artifacts present in EEG recordings, as well as from signals related to the actual neuronal activity. The instrumental artifacts may include, for example, residual MRI artifacts, EEG amplifier baseline drifts, signals due to mechanical vibrations, and interference signals picked up by EEG electrodes with poor electrical connections to the scalp. The physiological artifacts include eye blinking, residual cardioballistic signals, as well as muscle artifacts and all other artifacts present in conventional EEG. As mentioned in the introduction, ICA makes it possible to separate signals from different sources in multichannel EEG recordings. Using ICA, signals $V_i(t)$ from $N$ EEG channels can be approximated by linear combinations of $M$ non-Gaussian independent components (ICs) $F_j(t)$:

$$V_i(t) = \sum_{j=1}^{M} b_{ij} F_j(t) + \varepsilon_i(t), \quad i=1...N, M \leq N \qquad (6)$$

Here, $b_{ij}$ are elements of the matrix $\mathbf{W}^{-1}$ (the inverse of the unmixing matrix $\mathbf{W}$), describing projections of the found independent components $\{F_j\}$ back onto the EEG electrode space (e.g. Bell and Sejnowski, 1995; Makeig et al., 1997). The quantity $\varepsilon_i(t)$ is an error term also including Gaussian noise. In the present work, ICA is performed on the EEG data after the removal of both MRI and cardioballistic artifacts by means of the average artifact subtraction method. The ICA decomposition, Eq (6), makes it possible to select a subset $\{F_k\}$ of the independent components, consisting of those ICs ($k=1...K$) that correspond to EEG artifacts caused by random head motions, so that

$$V_{EMF}^{(i)}(t) \approx \sum_{k=1}^{K} b_{ik} F_k(t), \quad i=1...N, K \leq M \leq N \qquad (7)$$

The motion-related ICs are identified in the present study according to the following four criteria. *First*, the ICs of interest should reflect all major motion effects, observed in the EEG data, and exhibit high degrees of non-Gaussianity, as measured by the ICs' kurtosis or negentropy (Hyvärinen and Oja, 2000). The effects of rapid head movements are usually clearly visible both in the EEG recordings and in the IC time courses. In our studies, the ICs corresponding to rapid and random head movements typically have kurtosis values of the order of 10-1000, but ICs with lower kurtosis may be relevant as well. *Second*, the ICs should not be attributable to other known artifact sources such as eye blinking or residual cardioballistic effects. *Third*, the ICs of interest should have approximately bipolar topographies. An IC topography is a spatial map corresponding to a column of the $\mathbf{W}^{-1}$ matrix and describing projections of a given IC onto the EEG electrodes. A "bipolar" IC topography is defined here as the one that provides significant (i.e. exceeding a sufficiently high magnitude threshold) signal contributions, which exhibit opposite polarities for EEG channels on two opposite sides of the EEG array. Such bipolar topographies have routinely appeared in our ICA-based studies of both cardioballistic artifacts and EEG artifacts due to random head motions. We hypothesize that they reflect simple head rotations, i.e. rotations around a "fixed" axis passing through the head. The IC waveform in this case can be interpreted as time dependence of the angular speed of rotation (of unknown sign and amplitude) around such axis. Bipolar scalp topographies of EEG motion artifacts were mentioned in the work by Jansen et al., 2012, but those authors based their arguments on a physically unrealistic model of the head as a uniformly conductive sphere. *Fourth*, all the identified motion-related ICs should together provide a good approximation of the random motion artifacts when projected back onto the EEG electrode space, and removal of their projections from the EEG data should substantially reduce such artifacts.

The term $\Delta\Phi_D(t)$ in Eq (1) describes magnetic flux changes due to deformations of the effective contour. Such deformations result primarily from changes in pressure exerted on the EEG electrodes and their leads by the padding underneath and on both sides of the head during head movements. Unlike $\Delta\Phi_R(t)$, the term $\Delta\Phi_D(t)$ may depend on both rotational and translational motion parameters. The corresponding IC topographies may show contributions to EEG motion artifacts in parietal, occipital, and temporal regions. In the present study, we focus on the rotational effects, and neglect the deformational term $\Delta\Phi_D(t)$ for the sake of simplicity.

Comparison of Eqs (1), (5), and (7) suggests that time integrals of motion-related ICs can be used as motion regressors corresponding to different linear combinations of the rotational motion parameters. We use each IC from Eq (7) (with $k=1…K$, where $K$ is the total number of ICs describing random head motions) to generate two separate motion regressors as follows:

$$\int_0^t F_k(\tau)d\tau = \int_{t-\Delta t}^t F_k(\tau)d\tau + \int_0^{t-\Delta t} F_k(\tau)d\tau = R_1^{(k)}(t) + R_2^{(k)}(t) \quad (8)$$

Here, $\Delta t$ is a short constant time interval, and the integration limit $t-\Delta t$ is set to 0 for $t<\Delta t$. This definition ensures that short-term effects of rapid head movements are explicitly described by the regressors irrespective of the presence of long-lasting slow motions. Optionally, the ICs can be high-pass filtered prior to the integration to limit their effects to description of rapid motions. This procedure is equivalent to high-pass filtering of the EEG data, according to Eq (6). The IC-based regressors will not approximate any linear trends that may be present in the actual motion parameters $\varphi(t)$, $\psi(t)$, and $\theta(t)$, because any constant voltage offsets are removed by filtering during the EEG data acquisition, and all ICs have zero means. This is not a limitation for fMRI motion correction, however, because any linear trend in fMRI time series is explicitly modeled as a nuisance effect and removed during fMRI data analysis.

The motion regressors $R_1^{(k)}(t)$ and $R_2^{(k)}(t)$, defined in Eq (8), have the same temporal sampling as the EEG recordings. For correction of motion effects in fMRI data, these regressors need to be sub-sampled to match acquisition times $\{t_s\}$ for each slice in the fMRI dataset (after the steady state is reached). The correction of motion effects is then performed for each fMRI voxel's time series using the following linear regression procedure:

$$S_{fMRI}(t_s) = \beta_0 + \beta_1 R_L(t_s) + \sum_{k=1}^K [\beta_{k1} R_1^{(k)}(t_s) + \beta_{k2} R_2^{(k)}(t_s)] + \varepsilon(t_s) \quad (9)$$

Here, $\{\beta\}$ are fit coefficients for each 3D voxel, and $R_L(t_s)$ is a zero-mean linear regressor. The $R_1^{(k)}(t)$ and $R_2^{(k)}(t)$ regressors are linearly de-trended prior to their use in Eq (9). Subtraction of the fit terms containing $R_1^{(k)}(t)$ and $R_2^{(k)}(t)$ from the original fMRI time series for each voxel yields a motion-corrected fMRI dataset. The effects of such correction on fMRI data can be evaluated by comparing temporal signal-to-noise ratio (TSNR) values before and after the correction. The TSNR is defined as follows:

$$\text{TSNR} = \text{mean}(S_{fMRI}(t_s))/\text{std}(S_{fMRI}(t_s)) \quad (10)$$

It is an important characteristic of fMRI time courses, which depends on fMRI acquisition parameters, tissue type, and the amount of physiological noise (Bodurka et al., 2007).

The proposed method, E-REMCOR, makes it possible to generate motion regressors capable of describing rotational head motions with much finer, millisecond temporal resolution. It can improve efficiency of fMRI motion correction by providing a more accurate approximation of the effects of rapid head movements occurring on time scales shorter than the repetition time $TR$. It can be applied simultaneously with RETROICOR and other methods for fMRI physiological noise correction utilizing slice-time information.

*2.2 Experimental procedure*

The study was conducted at the Laureate Institute for Brain Research. The research protocol was approved by the Western Institutional Review Board (IRB). Five unmedicated MDD patients (mean age 32±11 years, three females) participated in the study. The patients are referred to throughout the paper as Subjects S1, S2, S3, S4, and S5. All the participants provided written informed consent as approved by the IRB. The experimental protocol included real-time fMRI neurofeedback training runs as well as resting fMRI runs (Zotev et al., 2011). EEG recordings were performed simultaneously with fMRI. Only resting-state EEG-fMRI results (one run per subject) are reported in this paper. For the resting run, the participants were instructed not to move, but to relax and rest while looking at the fixation cross on the screen. No subject reported any discomfort resulting from wearing an EEG cap during the experiment.

All functional and structural MR images were acquired using a General Electric Discovery MR750 whole-body 3 tesla MRI scanner with a standard 8-channel receive-only head coil array. A single-shot gradient-recalled EPI sequence with Sensitivity Encoding (SENSE, Pruessmann et al., 1999) was employed for fMRI. To enable accurate correction of MRI artifacts in EEG data, acquired simultaneously with fMRI, the EPI sequence was custom modified to ensure that the repetition time $TR$ was exactly 2000 ms (with 1 µs accuracy). The following EPI imaging parameters were used: FOV=240 mm, slice thickness=2.9 mm, slice gap=0.5 mm, 34 axial slices per volume, 96×96 acquisition matrix, echo time $TE$=30 ms, SENSE acceleration factor $R$=2, flip angle=90°, sampling bandwidth=250 kHz. The fMRI run time was 8 min 40 s. Three EPI volumes (6 s) were added at the beginning of the run to allow the fMRI signal to reach steady state, and were excluded from data analysis. The EPI images were reconstructed into a 128×128 matrix, so the resulting fMRI voxel size was 1.875×1.875×2.9 mm³. Physiological pulse oximetry and respiration waveforms were recorded (with 20 ms sampling interval) simultaneously with fMRI. A photoplethysmograph placed on the subject's finger was used for pulse oximetry, and a pneumatic respiration belt was used for respiration measurements. A T1-weighted magnetization-prepared rapid gradient-echo (MPRAGE) sequence with SENSE was used to provide an anatomical reference for the fMRI analysis. It had the following parameters: FOV=240 mm, 128 axial slices per slab, slice thickness=1.2 mm, 256×256 image matrix, $TR/TE$=5.0/1.9 ms, acceleration factor $R$=2, flip angle=10°, delay time $TD$=1400 ms, inversion time $TI$=725 ms, sampling bandwidth=31.2 kHz, scan time=4 min 58 sec.

The EEG recordings were performed simultaneously with fMRI using a 32-channel MR-compatible EEG system from Brain Products GmbH. Each subject wore an MR-compatible EEG cap (BrainCap MR from EASYCAP GmbH) throughout the experiment. The cap is fitted with 32 EEG electrodes (including Ref), arranged according to the international 10-20 system, and one ECG electrode placed on the subject's back. The EEG amplifier (BrainAmp MR plus) was positioned just outside the MRI scanner bore near the axis of the magnet approximately 1 m away from the subject's head. The amplifier was connected to the PC interface outside the scanner room via a fiber optic cable. The EEG system's clock was synchronized with the 10 MHz MRI scanner's clock using Brain Products' SyncBox device. The EEG signal acquisition was performed with 16-bit 5 kS/s sampling providing 0.2 ms temporal and 0.1 μV measurement resolution. The EEG signals measured relative to the standard reference (FCz as Ref) were hardware-filtered during the acquisition in the frequency band between 0.016 Hz (10 s time constant) and 250 Hz. The electrical cables connecting the EEG cap to the amplifier were fixed in place using sandbags. To reduce head motions, two foam pads were inserted in the MRI head coil on both sides of the subject's head. Consistency of the padding's firmness across multiple subjects could not be ensured, however. The EEG data acquisition was monitored in real time using Brain Products' RecView software, which enabled online correction of MRI and cardioballistic artifacts.

*2.3 Data analysis*

Processing of the EEG data, acquired simultaneously with fMRI, was performed using Brain Products' Analyzer 2 software. Removal of the MRI and cardioballistic artifacts was based on the average artifact subtraction method. The MRI artifact template was defined using the MRI slice markers. The slice markers were also used to select a 520-second-long EEG data interval, precisely corresponding to the fMRI time series of 260 volumes. After the MRI artifact removal, the EEG data were downsampled to 250 S/s sampling rate (4 ms sampling interval) and low-pass filtered at 40 Hz (48 dB/octave). The cardioballistic artifact template was determined from the cardiac waveform recorded by the ECG electrode, and the artifact to be subtracted was defined by a moving average over 21 cardiac periods.

Application of E-REMCOR includes three steps: i) independent component analysis of the EEG data; ii) integration of the components corresponding to major head motions; iii) correction of the fMRI dataset using the EEG-based motion regressors. In the present study, we used the FastICA algorithm (Hyvärinen, 1999), implemented in Analyzer 2. The ICA was applied to the EEG data from $N$=31 channels over the entire measurement time interval, and the number of ICs was set to $M$=31. The ICs corresponding to random head motions, Eq (7), were then identified as described above. The number $K$ of such ICs depends on the complexity of the subject's head movements, as well as performance of the ICA algorithm. In this work, we identified $K$=5 motion-related ICs for Subjects S1, S3, and S5, and $K$=4 ICs for Subjects S2 and S4. Time courses of the selected ICs were exported from Analyzer 2 and integrated in MATLAB to generate two motion regressors, $R_1^{(k)}(t)$ and $R_2^{(k)}(t)$, for each IC as defined in Eq (8), using $\Delta t$=0.4 s. These functions were linearly de-trended, and magnitudes of the resulting waveforms were scaled to fit the [−1,+1] interval.

In addition to E-REMCOR regressors defined in Eq (8), we considered an alternative set of regressors, which were based on the same motion-related ICs, $\{F_k(t)\}$, $k$=1…$K$, that were high-pass filtered at 0.1 Hz (time constant 1.6 s, 48 dB/octave) prior to the integration. Such regressors can describe the effects of rapid head movements without affecting any slow-motion variations in fMRI time courses.

Analysis of the fMRI data was performed in AFNI (Cox, 1996; Cox and Hyde, 1997). The linear regression procedure, Eq (9), was implemented using 3dTfitter program, which makes it possible to apply individual regressors to the time course of any voxel. On the image level, each regressor in Eq (9) was represented by an AFNI 3D+time dataset with an individual time course for each slice determined by sub-sampling the regressor's time series to match acquisition times for that slice in the original fMRI dataset. The least-squares solution of Eq (9) by means of 3dTfitter program yielded the coefficients $\{\beta\}$ as 3D datasets. E-REMCOR correction was performed by subtracting the terms, corresponding to the $R_1^{(k)}(t)$ and $R_2^{(k)}(t)$ regressors in Eq (9), from the original fMRI dataset. Motion effects in both the original and corrected fMRI data were further evaluated using 3dvolreg program in AFNI. This program performs slice timing correction and brain volume registration, and provides estimates of the six fMRI motion parameters. It also computes a maximum displacement in the brain automask for each fMRI volume and a root-mean-square (rms) difference between each volume and the registration base volume. The fMRI volume registration procedure was followed by a linear regression of six fMRI motion parameters and their first time derivatives using 3dDeconvolve program in AFNI. The general linear model (GLM) included 12 motion regressors and five polynomial terms in this case. The motion-related GLM terms were then subtracted from the volume registered dataset to yield a new dataset with the motion parameters regressed out.

To evaluate simultaneous performance of E-REMCOR and RETROICOR, we also carried out an fMRI data correction, in which terms with eight RETROICOR regressors were added to the right-hand side of Eq (9). They included four cardiac regressors, defined as $\cos(m\varphi_c)$ and $\sin(m\varphi_c)$ with $m=1,2$, and four respiratory regressors, defined as $\cos(m\varphi_r)$ and $\sin(m\varphi_r)$ with $m=1,2$. Here, $\varphi_c(t)$ is the cardiac phase and $\varphi_r(t)$ is the respiratory phase, determined from the physiological recordings simultaneous with fMRI (Glover et al., 2000). Similar to the E-REMCOR regressors, each RETROICOR regressor was represented by a 3D+time dataset with an individual time course for each slice obtained by the sub-sampling procedure based on the slice-time information.

The effects of E-REMCOR correction were evaluated by comparing TSNR values, Eq (10), for each 3D voxel's time series before and after the correction. For group analysis, the map of percent changes in TSNR after the application of E-REMCOR for each subject was transformed to the Talairach space (Talairach and Tournoux, 1988) using that subject's high-resolution anatomical brain image. The five single-subject maps were re-sampled to $2\times2\times2$ mm$^3$ isotropic voxel size, spatially smoothed using a Gaussian kernel with FWHM of 5 mm, and averaged. The same approach was used to evaluate group-level effects of the other motion correction steps as well.

## 3. Results

All five MDD patients exhibited significant head movements during the resting EEG-fMRI run. The maximum displacements in brain automasks over the entire run were found to be 2.2 mm, 1.8 mm, 1.6 mm, 1.9 mm, and 3.0 mm for subjects S1 through S5, respectively.

Figure 1A illustrates identification of the ICs reflecting major random head motions in the ICA decomposition of the EEG data for Subject S1. This subject exhibited significant movements during a time interval spanning several fMRI volumes, with the most drastic motion occurring during the second half of the data acquisition for volume 61. All fMRI volumes are numbered 1 through 260. The five ICs ($K$=5 out of total $M$=31) in Fig. 1A are highly non-Gaussian (kurtosis values 303, 242, 216, 26, and 26, respectively) and have bipolar topographies. Application of the same ICA procedure to the cardioballistic artifacts, subtracted previously from the EEG data using the average artifact subtraction method, also yielded several main cardioballistic ICs with bipolar topographies, but their kurtosis values were below 10. The topography of IC 1 in Fig. 1A is similar to that of the cardioballistic IC describing the pitch-like head rotation (i.e. rotation around the left-right axis) following the cardiac R peak. This similarity suggests that IC 1 also describes a pitch-like head rotation (the interhemispheric asymmetry of IC 1 indicates an uneven placement of the EEG cap and/or uneven positioning of the head within the scanner). Indeed, examination of the six fMRI motion parameters (shown in Fig. 4 below) reveals that two parameters have the largest peak values at volume 61: the pitch rotation ($\varphi=0.7$ deg) and the inferior-superior displacement ($z=-1.2$ mm). Therefore, the most drastic head motion during the acquisition of volume 61 is the motion through the EPI slice plane. Such through-plane motions are known to produce severe artifacts in fMRI data (Friston et al., 1996). The independent component IC 1 in Fig. 1A describes the pitch rotation as part of this motion. Motion-related ICs identified for the other four subjects (S2…S5) are exhibited in Supplementary Figs. 1 and 2.

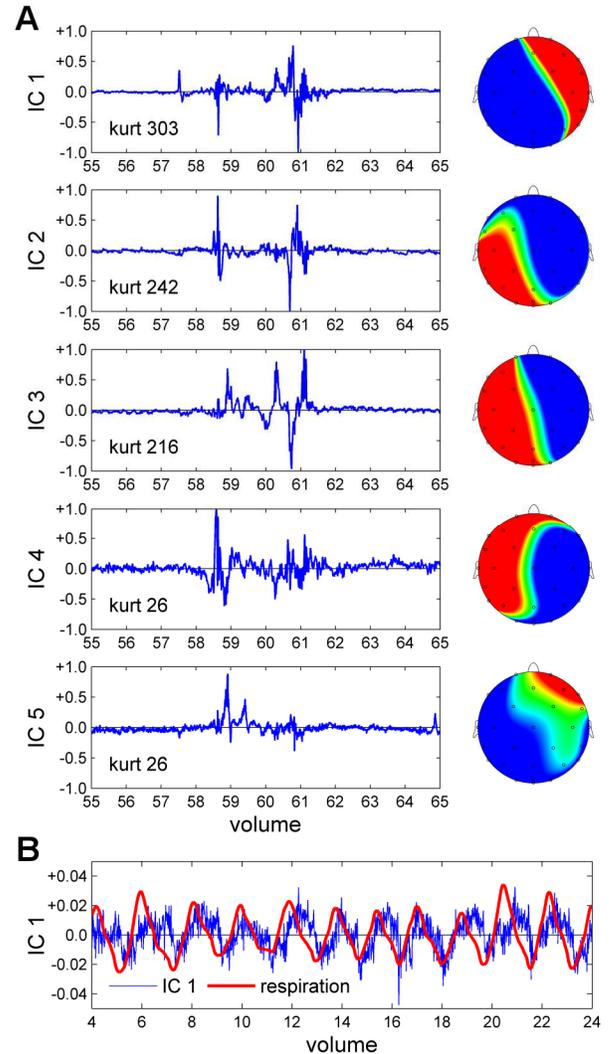

**Fig. 1.** A) EEG independent components (ICs) reflecting major random head motions of a single subject (Subject S1). The ICs' time courses (for a selected 20 s long interval, on the left) and the corresponding topographies (on the right) are shown. Temporal resolution is 4 ms. Each tick label along the time axis marks the end of data acquisition for a corresponding fMRI volume ($TR$=2 s). B) Comparison of the time course of IC 1 for an interval free of major head motions and the respiration waveform measured with 20 ms temporal resolution using a respiration belt.

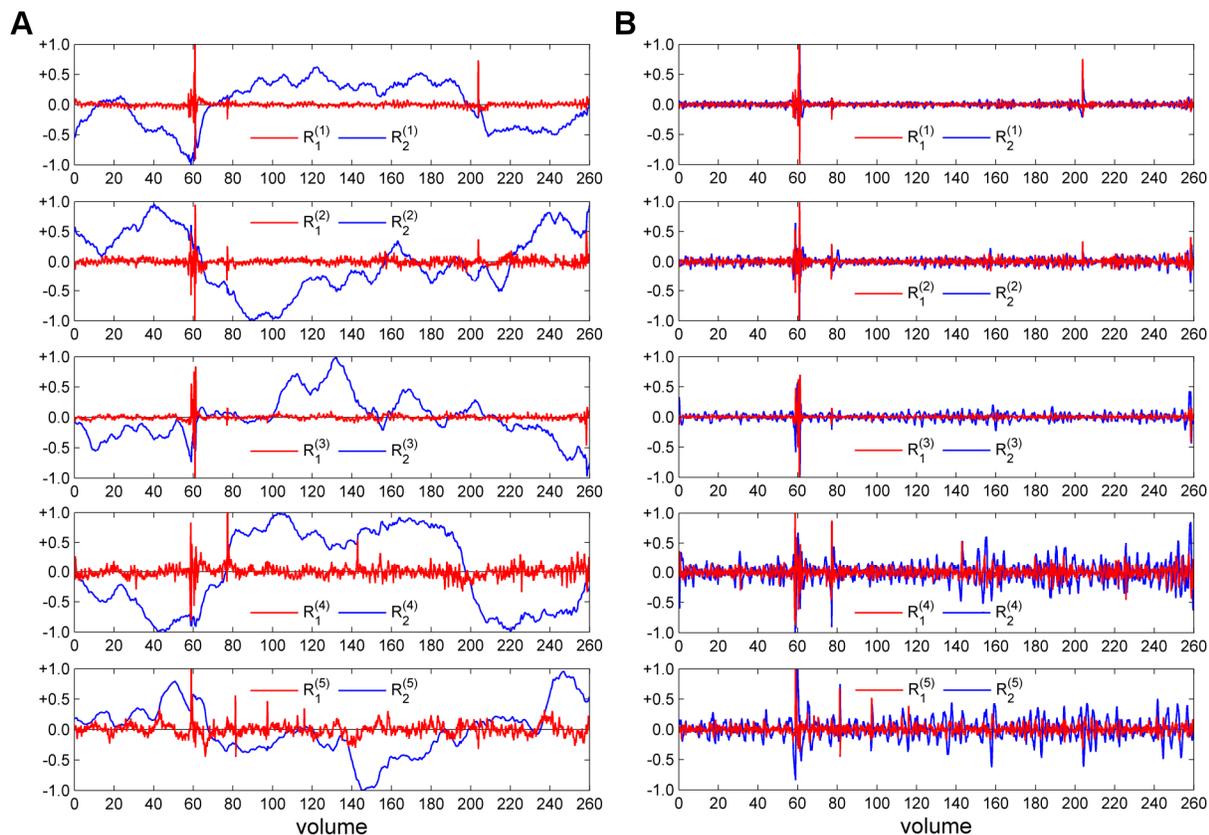

**Fig. 2.** A) Ten E-REMCOR motion regressors obtained by integration of time courses of the five EEG independent components in Fig. 1A (Subject S1). For each IC, numbered $k=1\ldots5$, regressor $R_1^{(k)}(t)$ is depicted in red and regressor $R_2^{(k)}(t)$ is shown in blue. Temporal resolution is 4 ms. Tick labels along the time axis mark fMRI volumes acquired concurrently with EEG. B) Ten E-REMCOR regressors obtained by integration of the five ICs (Fig. 1A) that were high-pass filtered at 0.1 Hz.

Figure 1B compares the time course of IC 1 to the respiration waveform, which was measured simultaneously with EEG-fMRI using the pneumatic respiration belt. Close correspondence between the two time courses (up to the IC's unknown sign and scale) is observed across the entire run. This means that the EEG array consistently registered head motions due to respiration in addition to random head movements and cardioballistic motions.

Figure 2A exhibits time courses of ten E-REMCOR regressors ($R_1^{(k)}(t)$ in red, $R_2^{(k)}(t)$ in blue, $k=1\ldots5$) for Subject S1 obtained by time integration of the five ICs in Fig. 1A. The integration was performed as described in the Methods section, and the regressors were scaled to fit the $[-1,+1]$ interval. Figure 2B shows an alternative set of regressors based on the same motion-related ICs (Fig. 1A), that were high-pass filtered at 0.1 Hz (see Methods). Note that all results reported below were obtained using E-REMCOR regressors based on unfiltered motion ICs as in Fig. 2A, unless explicitly stated otherwise.

Results of the E-REMCOR correction of the original fMRI data for Subject S1 are shown in Fig. 3A. The correction was performed as described in the Methods section. All acquired axial EPI slices are numbered 1 through 34 in the inferior-superior direction, and slices 2…33 are shown in Fig. 3A. The spacing between central planes of adjacent EPI slices is 3.4 mm. The results in Fig. 3A demonstrate substantial improvements in TSNR after the application of E-REMCOR. The maximum TSNR

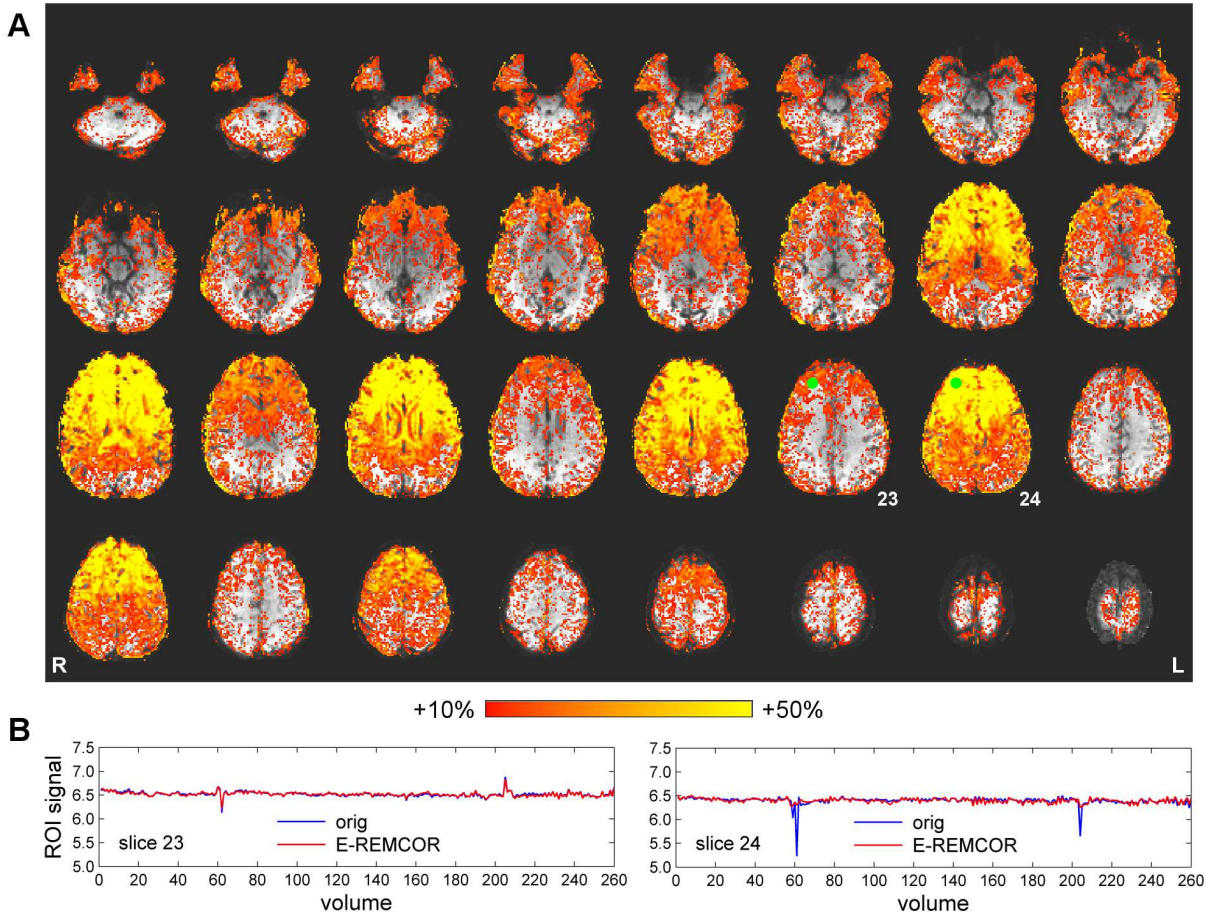

**Fig. 3.** A) Improvements in TSNR of single-subject fMRI data (Subject S1) after the application of E-REMCOR with ten motion regressors exhibited in Fig. 2A. The montage shows 32 axial EPI slices numbered 2…33. The underlay image is a TSNR map for the original fMRI data (greyscale, display range 0−150), and the overlay (color) image is a map of the percent change in TSNR as a result of the E-REMCOR correction. B) Time courses of two 12 mm diameter single-slice ROIs, defined within EPI slices 23 and 24 and marked by green circles in Fig. 3A, before and after the E-REMCOR procedure (Subject S1).

change is 138%, and a large number of voxels show TSNR increases by 50% or more.

One interesting feature of the results in Fig. 3A is that the effects of E-REMCOR are significantly stronger for the even-numbered slices than for the odd-numbered slices. This is a consequence of the fact that the original fMRI data for Subject S1 exhibited reduced TSNR levels for the even-numbered slices compared to the odd-numbered slices. The reason for this is the following. During an interleaved EPI acquisition, the first half of each $TR$ interval is used to acquire all odd-numbered slices, and the second half – to acquire all even-numbered slices. Because the largest through-plane motion occurred during the second half of the $TR$ interval for volume 61 (Fig. 1A), it affected the even-numbered slices, causing large fMRI signal variations and increasing standard deviations of the time courses. This effect is illustrated even further in Fig. 3B, which compares time courses of two identical 12 mm diameter single-slice ROIs defined one above the other within two adjacent EPI slices in the frontal brain region (23 and 24 in Fig. 3A). Between the two time courses, the one for slice 24 shows the larger signal variation in the original fMRI data and the stronger effect of the E-REMCOR correction.

Figure 4A compares time courses of the six fMRI motion parameters for Subject S1 before and after the E-REMCOR correction using the regressors in Fig. 2A. The motion parameters were determined from the volume registration of all fMRI volumes in a given dataset to the 1st volume as a registration base. According to Fig. 4A, the motion parameters for the E-REMCOR-corrected data are generally similar (though not identical) to those for the original fMRI data, but magnitudes of their peaks, corresponding to rapid head movements, are substantially reduced. Figure 4B exhibits motion parameters for the same subject after the E-REMCOR correction using the regressors in Fig. 2B. Because these regressors do not affect slow-motion-related fMRI signal variations, the motion parameters for the E-REMCOR-corrected data in Fig. 4B are almost identical to the original motion parameters, but the effects of rapid head movements are still efficiently reduced.

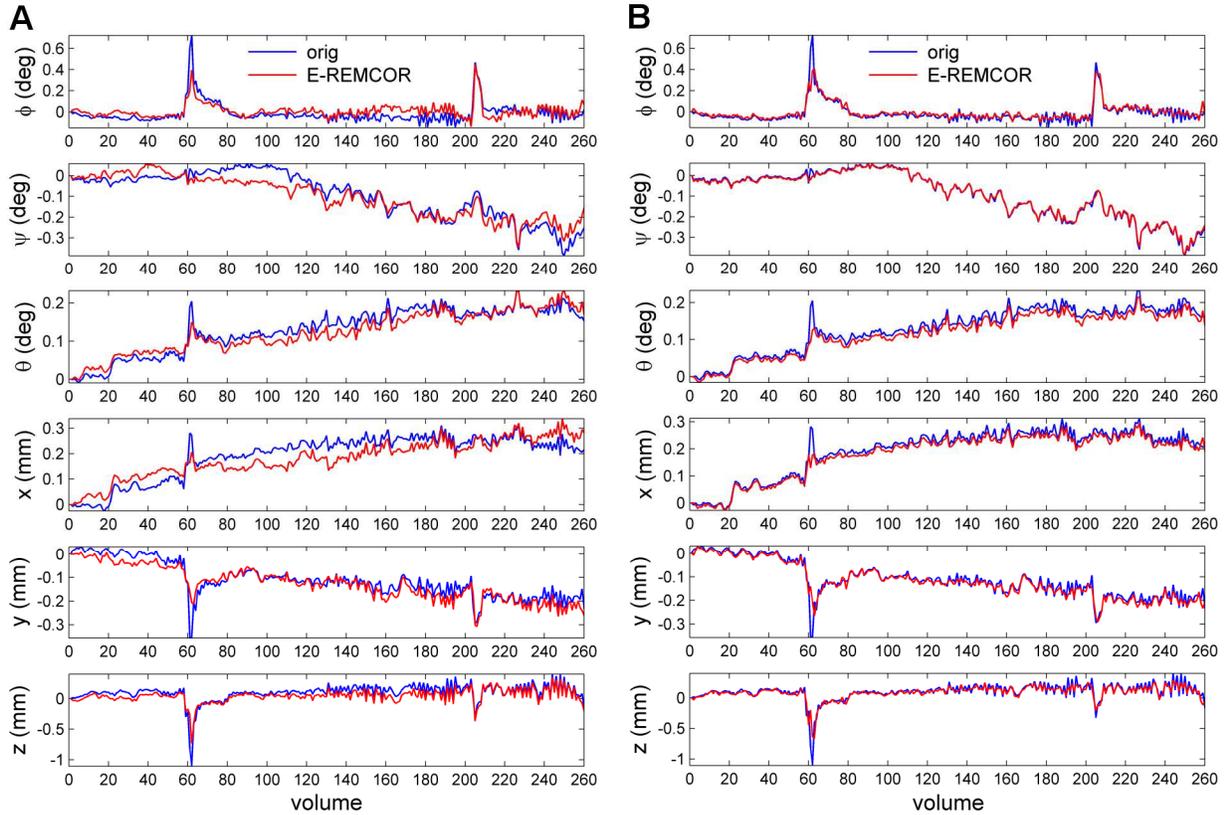

**Fig. 4.** Time courses of the six fMRI motion parameters before and after the application of E-REMCOR (Subject S1). A) Using E-REMCOR regressors in Fig. 2A. B) Using E-REMCOR regressors in Fig. 2B.

Performance of E-REMCOR for each of the five subjects is illustrated in Fig. 5A. Four axial EPI slices with 10.2 mm spacing are shown for each subject. The lowest slice in each case is located approximately 10 mm above the superior edge of the anterior commissure. Figure 5A demonstrates that E-REMCOR visibly improves TSNR for all the subjects. The images in Fig. 5A show TSNR improvements both for voxels near the edges of the brain and for broad inner brain regions. In contrast, fMRI volume registration leads to substantial TSNR improvements near the edges of the brain only, according to Supplementary Fig. 3. Figure 5B exhibits the difference in TSNR between the E-REMCOR-corrected data and the original fMRI data (for each subject) after both datasets were volume registered to the same base – the 1st volume of the original fMRI data. Note that the TSNR differences in Fig. 5B are predominantly positive across the brain.

Figure 6A exhibits the maximum displacement in the brain automask before and after the E-REMCOR correction for each of the five subjects. According to Fig. 6A, the application of E-REMCOR substantially reduces the maximum displacement for the volumes affected by rapid head movements. This effect is similar to the one observed in Fig. 4A for the motion parameters. Figure 6B shows the rms difference between a given fMRI volume and the base volume after the volume registration. The rms difference is a direct measure of similarity between two fMRI volumes, and it is the quantity minimized by the volume registration procedure. According to Fig. 6B, the rms difference is substantially reduced after the E-REMCOR correction for the volumes affected by rapid head movements. Importantly, this difference is lower in the E-REMCOR-corrected data after volume registation than in the original fMRI data after volume registration for almost all fMRI volumes across the five subjects (Fig. 6B).

To further evaluate the effects of E-REMCOR, we carried out a direct volume-by-volume comparison of the E-REMCOR-corrected data after volume registration and the original fMRI data after volume registration. Both the original dataset and the corrected one were volume registered to the same base – the 1st volume of the original fMRI data. Maximum displacements in brain automasks between the corresponding volumes of the two datasets were then estimated by means of an individual volume registration for each pair of volumes (i.e. 1st to 1st, 2nd to 2nd, 3rd to 3rd, and so on). The estimated maximum displacements are shown in Fig. 7 for each of the five subjects. As expected, the largest maximum displacements are observed for the volumes affected by rapid head movements (compare to Fig. 6), which are the primary targets for the E-REMCOR correction. For all other volumes, the maximum displacements are of the order of 0.05 mm or less, according to Fig. 7.

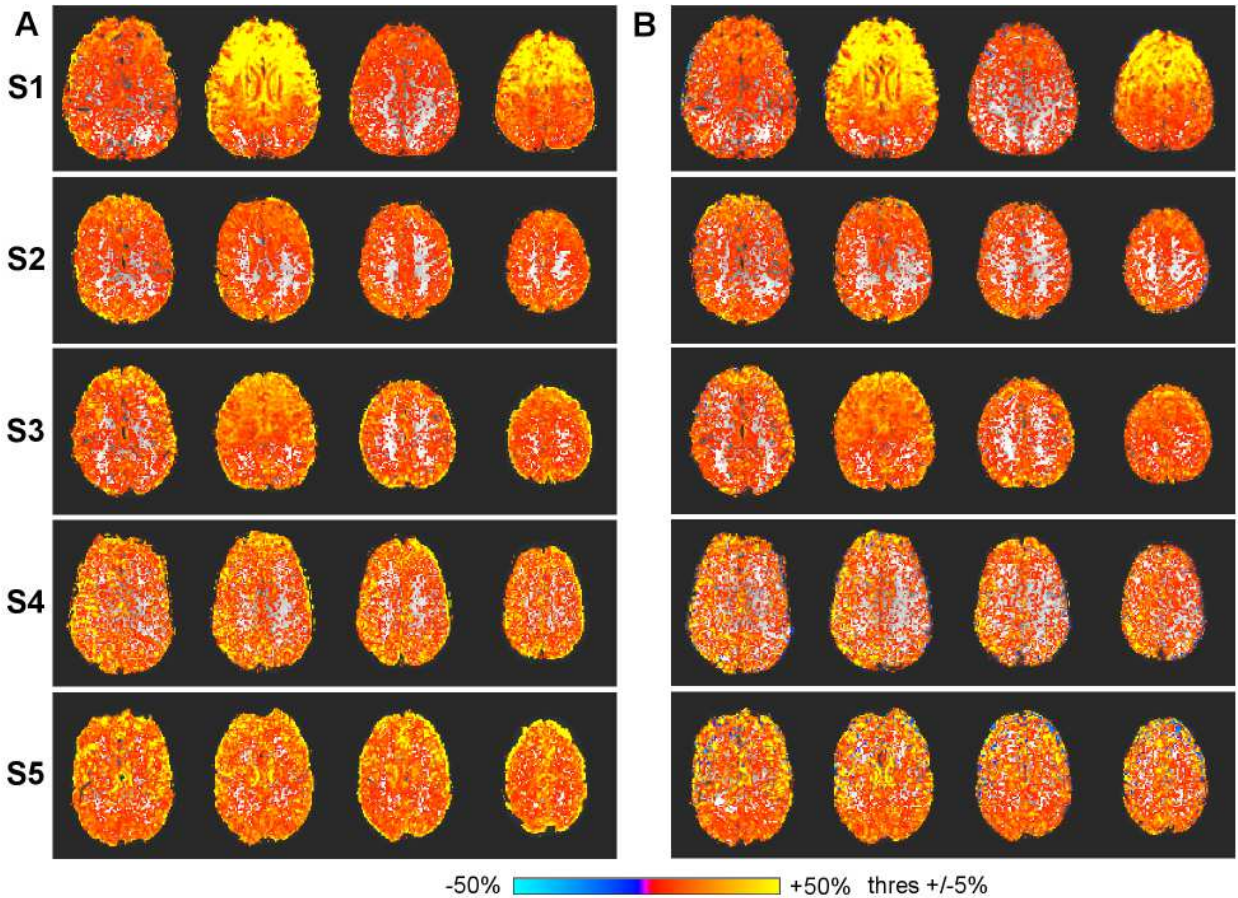

**Fig. 5.** A) Improvements in TSNR after the application of E-REMCOR for five subjects (S1…S5). B) Differences in TSNR between the E-REMCOR-corrected data after volume registration and the original fMRI data after volume registration. Four axial EPI slices with 10.2 mm spacing are shown for each subject.

Average results of the E-REMCOR correction for the group of five subjects are shown in Fig. 8. The TSNR percent change maps were transformed to the Talairach space, processed, and averaged as described in the Methods section. The axial slices in Fig. 8 (A,B,C,D) have 5 mm spacing, with bottom/top slices corresponding to $z=-20$ mm and $z=55$ mm, respectively.

The results in Figs. 8A and 8B were obtained using E-REMCOR regressors based on unfiltered motion ICs as in Fig. 2A. The largest average TSNR improvement after the application of E-REMCOR in Fig. 8A is 37%, and many areas exhibit TSNR enhancements as high as 25%. Similar to the single-subject results in Fig. 5A, the average results show substantial TSNR increases near the surface of the brain, including the medial plane. The effects of E-REMCOR are most pronounced in the frontal brain areas, which typically exhibit the largest through-plane movements. Importantly, the TSNR improvements in Fig. 8A are statistically significant ($p<0.05$) for most brain voxels, as demonstrated in Supplementary Fig. 4. Figure 8B shows the average difference in TSNR between the E-REMCOR-corrected data after volume registration and the original fMRI data after volume registration for the five subjects. The largest TSNR enhancement in Fig. 8B is 31%. The results in Fig. 8B demonstrate that TSNR improvements attained with E-REMCOR are largely preserved after volume registration.

The results in Figs. 8C and 8D were acquired using E-REMCOR regressors based on the motion ICs high-pass filtered at 0.1 Hz as in Fig. 2B (See Methods). The maximum average TSNR improvement after the application of E-REMCOR in Fig. 8C is 18%, and large frontal brain regions show TSNR increases by as much as 15%. These values are lower than in Fig. 8A, because the regressors used (Fig. 2B) approximate the effects of rapid head movements without affecting fMRI signal variations caused by slow motions (Fig. 4B). As a result, fewer fMRI volumes are affected by the E-REMCOR correction in this case, but TSNR is still computed across all volumes in the fMRI dataset. Figure 8D shows the average difference in TSNR between the E-REMCOR-corrected data after volume registration and the original fMRI data after volume registration. The map in Fig. 8D is very similar to that in Fig. 8C. These results suggest that rapid head movements have the strongest effects on fMRI time courses for the frontal brain regions.

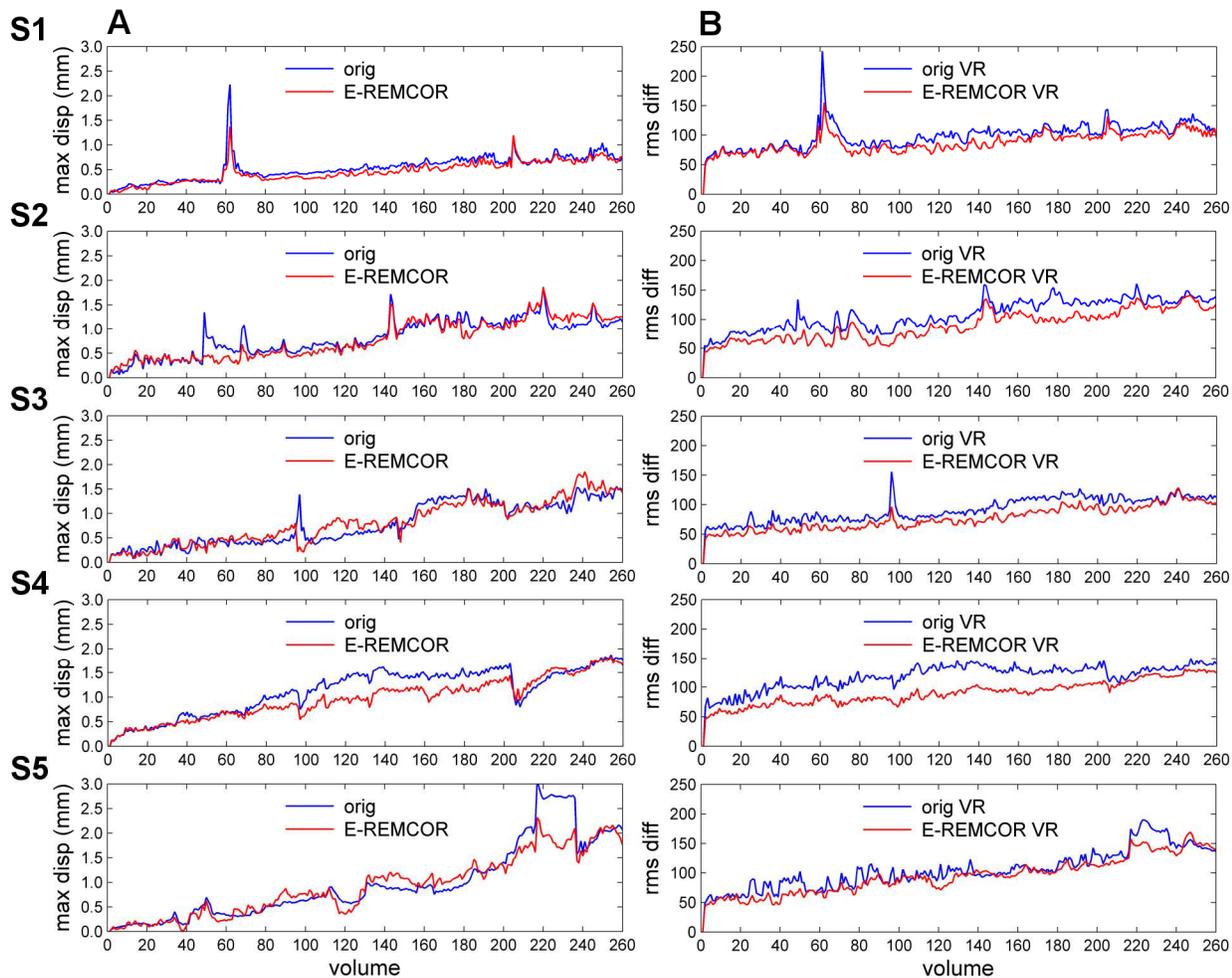

**Fig. 6.** A) Maximum displacements in brain automasks, estimated via volume registration with the 1st volume, for the fMRI datasets before and after the E-REMCOR procedure for five subjects (S1…S5). B) Root-mean-square differences (in image intensity units) between a given fMRI volume and the 1st volume for the original fMRI data after volume registration (VR) and for the E-REMCOR-corrected data after volume registration.

Figure 9 compares average results of the fMRI data correction by RETROICOR (Fig. 9A) and those after the simultaneous correction by E-REMCOR and RETROICOR (Fig. 9B). The simultaneous linear regression was performed as described in the Methods section. Comparison of the results in Fig. 9A and in Fig. 8A shows that E-REMCOR and RETROICOR complement each other: while E-REMCOR provides TSNR improvements in the frontal brain regions and near the surface of the brain, RETROICOR leads to TSNR enhancements in the areas close to blood vessels, such as sulci and the regions near the brain stem. Figure 9B demonstrates that the simultaneous application of E-REMCOR and RETROICOR improves TSNR values across almost the entire brain by at least 10% over the original fMRI data. The maximum TSNR enhancement is 43% in Fig. 9B, compared to 25% in Fig. 9A. Overall TSNR improvements by as much as 35% are observed for many brain regions in Fig. 9B. Comparison, after summation of the corresponding maps in Fig. 8A and Fig. 9A, with the map in Fig. 9B shows additional TSNR increases by up to 6% specifically due to the simultaneous regression, both in the regions affected by RETROICOR and in those affected by E-REMCOR.

Figure 10 demonstrates the benefits of E-REMCOR for fMRI motion correction, when the traditional processing with both volume registration and regression of fMRI motion parameters is employed. The regression of six fMRI motion parameters and their first time derivatives was carried out as described in the Methods section. Figure 10A shows the average TSNR improvement over the original (unprocessed) fMRI data after the volume registration and regression of motion parameters. The maximum TSNR increase in Fig. 10A is 102%. However, the large motion correction effects are only observed near the outer surface of the brain. This correction pattern is typical of the volume registration procedure (see Supplementary Fig. 3). The average TSNR enhancement over the original fMRI data after the E-REMCOR correction, followed by the volume registration and regression of motion parameters, is exhibited in Fig. 10B. The largest TSNR improvement in Fig. 10B is 115%, and the correction pattern is more uniform across the brain. Note that the sets of motion parameters determined from the volume registration procedure are somewhat different for the original and the E-REMCOR-corrected data (see Fig. 4A).

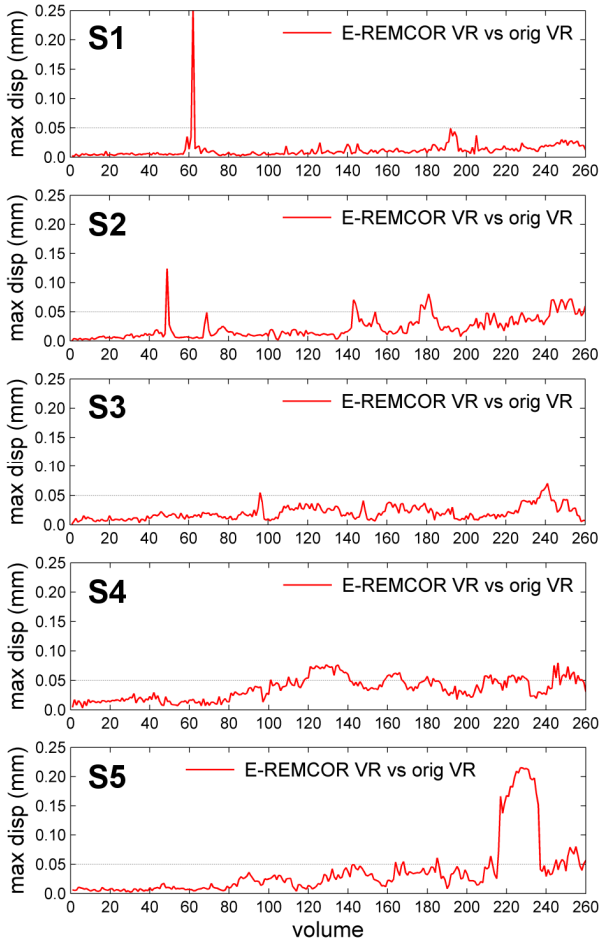

**Fig. 7.** Maximum displacements in brain automasks, estimated via individual registration of the corresponding volumes (1st to 1st, 2nd to 2nd, etc) between the E-REMCOR-corrected data after volume registration (VR) and the original fMRI data after volume registration.

Figure 10C shows the average difference in TSNR between the E-REMCOR-corrected data after volume registration and regression of motion parameters and the original fMRI data after volume registration and regression of motion parameters. The largest TSNR difference in Fig. 10C is 30%, and many brain regions show TSNR improvements by as much as 20%. Similarly, Figure 10D exhibits the average difference in TSNR between the fMRI data, corrected simultaneously by E-REMCOR and RETROICOR with the subsequent volume registration and regression of motion parameters, and the original fMRI data after volume registration and regression of motion parameters. The maximum TSNR difference in Fig. 10D is 42%, and TSNR improvements by as much as 30% are observed for many brain areas. Note that the average TSNR differences in Figs. 10C and 10D are uniformly positive across the brain.

To perform a quantitative comparison of the motion correction results in Fig. 8A, Fig. 10A, and Fig. 10B, we computed mean % TSNR improvements for ten ROIs corresponding to ten bilateral Brodmann areas (BA) in the frontal part of the brain. The ROIs were defined using the Talairach-Tournoux atlas in AFNI. The following results were obtained for the data in Fig. 8A, Fig. 10A, and Fig. 10B, respectively. BA 6: 15.1%, 31.9%, 43.3%; BA8: 17.2%, 39.1%, 51.5%; BA9: 19.6%, 37.3%, 53.6%; BA10: 21.6%, 32.7%, 51.4%; BA24: 15.9%, 20.7%, 32.6%; BA32: 19.0%, 24.7%, 40.2%; BA44: 21.5%, 28.4%, 49.7%; BA45: 20.1%, 34.3%, 52.7%; BA46: 19.0%, 37.2%, 55.0%; BA47: 15.8%, 27.0%, 40.0%. The mean TSNR improvements over the original fMRI data across the ten areas are 18.5% (±2.4% SD), 31.3% (±6.0%), and 47.0% (±7.5%), respectively, for the E-REMCOR correction alone (Fig. 8A), for the traditional volume registration with regression of motion parameters (Fig. 10A), and for the E-REMCOR correction followed by the volume registration with regression of motion parameters (Fig. 10B). These results are further examined in the Discussion section below.

The effects of E-REMCOR correction on single-subject resting-state functional connectivity results are illustrated in Supplementary Figs. 5 and 6.

## 4. Discussion

We have developed a novel EEG-assisted method for retrospective motion correction of fMRI data. E-REMCOR takes advantage of the ability of the EEG sensor array to detect rotational head motions inside an MRI scanner in real time with millisecond temporal resolution. This ability is illustrated in Fig. 1A and Supplementary Figs. 1 and 2. The EEG array as a head motion detector is particularly sensitive to rapid head movements. It is also sensitive enough to register small and slow head motions, such as the motions due to respiration, as demonstrated in Fig. 1B. Another important conclusion one can draw from Fig. 1B is that motion-related ICs, identified in this work by their ability to describe large and rapid head movements, also contain

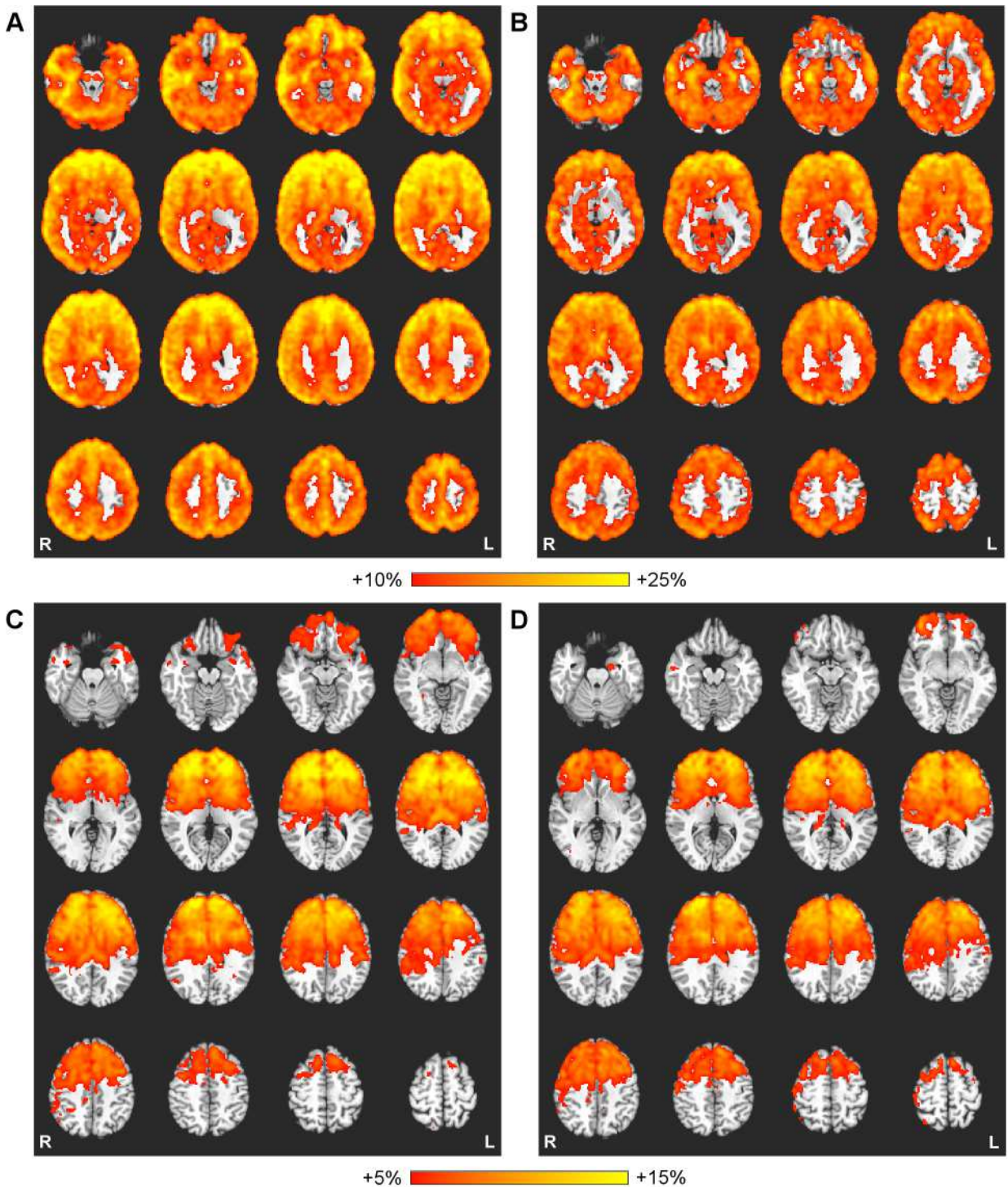

**Fig. 8.** Group results. A) Average improvement in TSNR for five subjects after the application of E-REMCOR with regressors based on unfiltered motion ICs (as in Fig. 2A). B) Average difference in TSNR between the E-REMCOR-corrected data (Fig. 8A) after volume registration and the original fMRI data after volume registration. C) Average improvement in TSNR for five subjects after the application of E-REMCOR with regressors based on motion ICs high-pass filtered at 0.1 Hz (as in Fig. 2B). D) Average difference in TSNR between the E-REMCOR-corrected data (Fig. 8C) after volume registration and the original fMRI data after volume registration. All the results are projected onto the standard anatomical template (TT_N27) in the Talairach space and shown with 5 mm spacing between the axial slices.

information about small and slow head motions.

We demonstrated in the Methods section that the ICs describing rigid-body head rotations in the ICA decomposition of EEG data, recorded inside an MRI scanner, are related to rotational head motion parameters (Eqs (1), (5), (7)). Because the ICs are defined statistically rather than analytically, a rigorous definition of a complete basis set is not applicable. Consequently, the precise number of motion-related ICs may be unknown, and their physical interpretation may be non-trivial. However, an important advantage of E-REMCOR regressors, in addition to their high temporal resolution, is that they are individually tailored to describe each subject's independent rotational head movements as detected by EEG and separated by ICA. The number and properties of E-REMCOR regressors will depend on the complexity of the subject's movements (statistical independence of the ICs ensures that their time courses are uncorrelated). fMRI motion parameters, in contrast, describe the overall head motion, which is a superposition of the independent movements. As discussed in the Introduction, fMRI signal variations due to motions can be caused not only by the actual rigid-body head displacements, but also by the concomitant changes in spin history and spatial distributions of magnetic susceptibility artifacts. These effects can conceivably be somewhat different for different independent movements contributing to the overall head motion. Therefore, E-REMCOR regressors can potentially provide more flexibility in approximating the effects of rotational head motions on fMRI time courses. Furthermore, ICA does not require a continuous input waveform and can be applied to EEG data from a number of separate intervals. This property makes it possible to generate regressors describing specific head motions of interest (for example, those due to occasional swallowing), provided that time intervals containing such motions can be reliably identified in the EEG recordings. Unlike fMRI motion parameters, determined from the fMRI volume registration procedure, E-REMCOR regressors are completely independent of the fMRI data and any artifacts that may be present in those data.

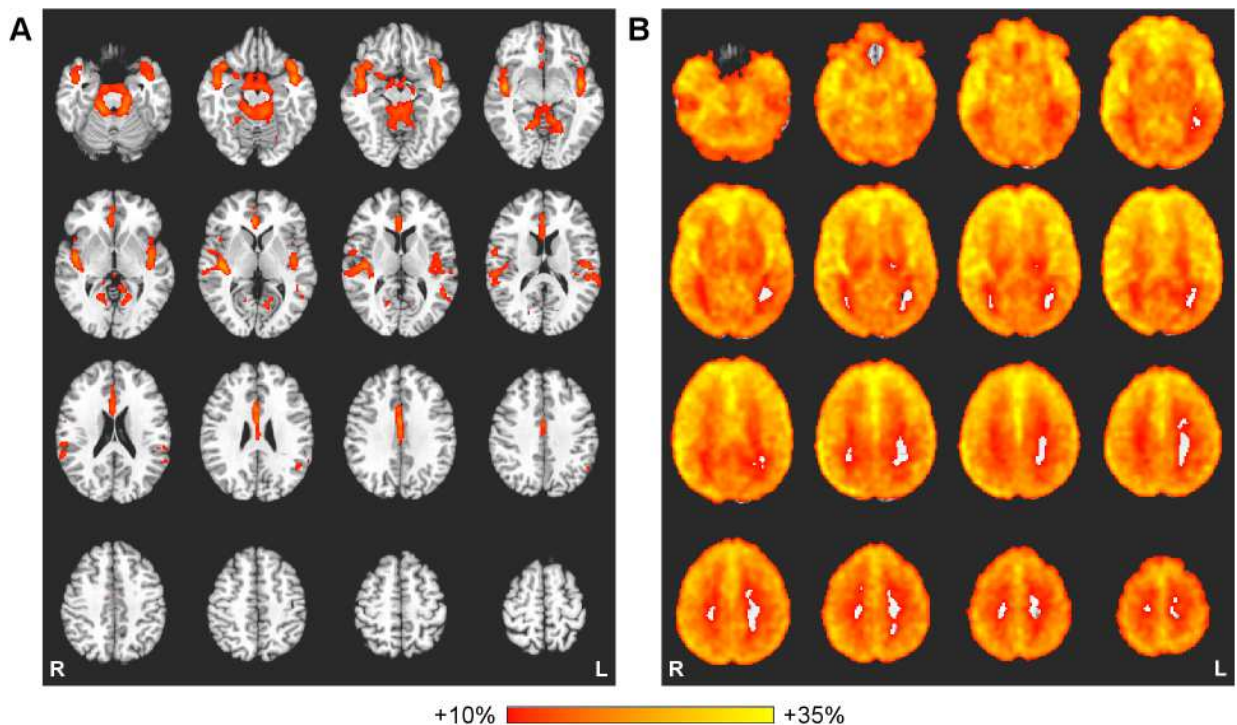

**Fig. 9.** Group results. A) Average improvement in TSNR for five subjects after the application of RETROICOR with 8 regressors. B) Average improvement in TSNR for five subjects after the simultaneous application of E-REMCOR (8-10 regressors based on unfiltered motion ICs) and RETROICOR (8 regressors). The results are shown in the Talairach space with 5 mm spacing between the slices.

E-REMCOR is a pre-processing technique to be applied to the original fMRI data with unaltered slice-time properties prior to the slice-timing correction, volume registration, and regression of fMRI motion parameters. From this point of view, E-REMCOR is similar to RETROICOR. The main purpose of E-REMCOR is to take advantage of the high temporal resolution of EEG to improve correction of the effects of rapid head movements, occurring on time scales shorter than *TR*. As demonstrated in Fig. 3, such movements may strongly affect just a few slices within an fMRI volume, making the motion appear non-rigid and the affected fMRI volume – non-uniform. An attempt to align such a motion-distorted volume with the base volume will result in a substantial volume registration error (Fig. 6B) and produce incorrect values of the motion parameters. Thus, the performance of the volume registration procedure can be greatly impaired for those fMRI volumes that are affected by rapid head movements. Correction of such rapid-motion effects by means of E-REMCOR reduces the motion-related intra-volume signal variance (Fig. 3B), making the volume more uniform and enabling a more accurate volume registration (Fig. 6B). The fMRI motion parameters, determined from the volume registration of the E-REMCOR-corrected data, show substantial reduction in magnitudes of the peaks corresponding to such rapid-motion-affected volumes (Fig. 4A,B). The maximum displacement in brain automask exhibits a similar behavior (Fig. 6A).

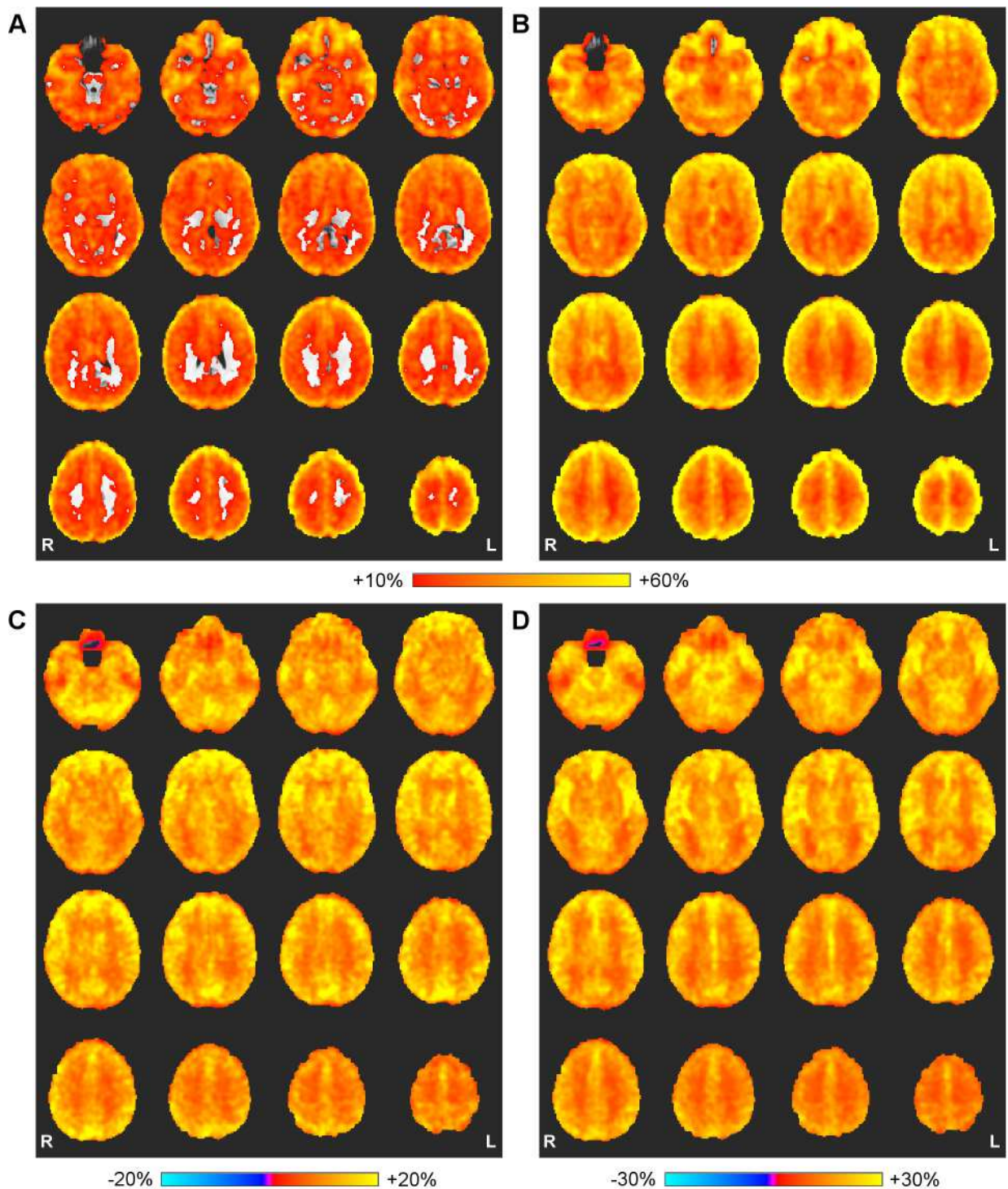

**Fig. 10.** Group results. A) Average improvement in TSNR for five subjects after volume registration and regression of six motion parameters (and their time derivatives). B) Average improvement in TSNR after E-REMCOR correction, volume registration, and regression of motion parameters (and their time derivatives). C) Average difference in TSNR between the E-REMCOR-corrected data after volume registration and regression of motion parameters (and their time derivatives) and the original fMRI data after volume registration and regression of motion parameters (and their time derivatives). D) Average difference in TSNR between the fMRI data, corrected using simultaneus E-REMCOR and RETROICOR (as in Fig. 9B) with the subsequent volume registration and regression of motion parameters (and their time derivatives), and the original fMRI data after volume registration and regression of motion parameters (and their time derivatives). The results are shown in the Talairach space with 5 mm spacing between the slices.

It is also important to consider the effects of E-REMCOR on the performance of the volume registration procedure for those fMRI volumes that are *not* affected by rapid head movements. Our results indicate that such effects are insignificant. *First*, TSNR values for the E-REMCOR-corrected data after volume registration are higher than for the original fMRI data after volume registration for most of the voxels (Fig. 5B). This suggests that E-REMCOR does not interfere with the performance of the subsequent volume registration procedure. *Second*, the rms difference between a given fMRI volume and the base volume after the volume registration is lower for the E-REMCOR-corrected data than for the original fMRI data for most of the volumes (Fig. 6B). This indicates that E-REMCOR makes fMRI volumes more similar to the base volume and thus improves the volume registration. *Third*, the maximum volume-by-volume displacements (Fig. 7) between the E-REMCOR-corrected data after volume registration and the original fMRI data after volume registration are of the order of 0.05 mm for any volumes not affected by rapid head movements. These displacements are much smaller than the fMRI voxel size in our experiments ($1.875 \times 1.875 \times 2.9$ mm$^3$). This means that E-REMCOR does not cause any substantial misalignments of voxels in the volume registered dataset that could affect the detection capability for fMRI activation. Taken together, these results indicate that E-REMCOR generally complements the volume registration procedure without reducing its performance. Clearly, the volume registered versions of the original and E-REMCOR-corrected data should always be compared to ensure their consistency across voxels (Fig. 5B) and across volumes (Fig. 6B). Also, it is always possible to use the E-REMCOR regressors based on high-pass filtered motion ICs (Fig. 2B) that only correct the effects of rapid head movements, while leaving any slow-motion effects unchanged (Fig. 4B). One should keep in mind, however, that the low-frequency signal variations in the motion ICs' time courses contain important information about slow head motions, as demonstrated in Fig. 1B.

Performance of E-REMCOR in terms of TSNR enhancement is demonstrated in Figs. 3, 5, 8, 9, and 10. Application of E-REMCOR to the original (unprocessed) fMRI data leads to TSNR improvements as high as 50% in single-subject analysis (Fig. 5A) and as high as 25% after group averaging (Fig. 8A). An important result of this work is that the TSNR improvements achieved with E-REMCOR persist through all the stages of the traditional processing. When both the E-REMCOR-corrected data and the original fMRI data are subjected to volume registration, the average TSNR differences for the motion-affected brain regions are around 20% (Fig. 8B). When the resulting datasets are subjected to regression of fMRI motion parameters, the average TSNR differences are still around 20% (Fig. 10C). This means that the TSNR effects of E-REMCOR cannot be achieved or approximated with the traditional processing. The reason is that E-REMCOR provides a *slice-specific* motion correction (Eq (9)), while the volume registration procedure and regression of motion parameters operate on volume-by-volume basis. The slice-specific correction by E-REMCOR reduces the motion-related *intra-volume* variance (Fig. 3) prior to the other processing steps. This leads to substantial TSNR improvements (Fig. 5A), and also benefits the volume registration (Fig. 6B) and regression of motion parameters by making fMRI volumes more uniform. While these effects are most pronounced for fMRI volumes affected by rapid head movements, they are also observed for volumes affected by slower motions (Fig. 6B).

Because TSNR improvements vary from voxel to voxel, a quantitative comparison of different correction approaches requires an ROI selection, and depends on this selection. For the set of ROIs in the frontal part of the brain, considered in the Results section, the E-REMCOR correction provided mean group-level TSNR improvement over the original fMRI data by 18.5%, the traditional volume registration with regression of motion parameters – by 31.3%, and the combination of E-REMCOR with the traditional volume registration and regression of motion parameters – by 47.0%. While the E-REMCOR correction was less efficient than the traditional processing (18.5% vs 31.3%), their combination led to an additional 15.7% average increase in TSNR of the resulting fMRI data (47.0% vs 31.3%). This additional TSNR increase is only slightly lower than the 18.5% TSNR improvement by E-REMCOR alone, suggesting that the two approaches complement each other. Moreover, this TSNR increase constitutes a 50% improvement in motion correction efficiency over the traditional volume registration and regression of motion parameters. This remarkable and somewhat surprising result demonstrates that E-REMCOR as a pre-processing technique can substantially enhance fMRI motion correction.

An attractive property of E-REMCOR is that it can be applied simultaneously with RETROICOR and other methods for physiological noise correction, as demonstrated in Fig. 9B. Such simultaneous regression improves the quality of both the random head motion correction by E-REMCOR and the physiological noise correction by RETROICOR (see Results). This observation agrees with the conclusion, reached previously in the work by Jones et al., 2008, that motion correction improves RETROICOR performance. The simultaneous application of E-REMCOR and RETROICOR leads to TSNR improvements by as much as 35% (Fig. 9B), and these effects are largely preserved after the volume registration and regression of motion parameters (Fig. 10D). In principle, it is possible to use the ICs corresponding to cardioballistic artifacts in EEG data to generate regressors describing the small rigid-body head rotations due to cardiac activity. Such regressors can be somewhat redundant if RETROICOR is used at the same time. However, if random-motion and cardioballistic artifacts in EEG data are hard to separate, all motion-related ICs should be used to define E-REMCOR regressors.

A limitation of E-REMCOR reflects the fact that EEG motion artifacts are caused primarily by rotational head movements. Depending on the orientation of fMRI slices and the definition of the image reference frame, the artifacts will, in general, be functions of either three (pitch, roll, yaw) or two (pitch, roll) rotational motion parameters. The artifacts will not reflect translational head movements, if there are no deformations of the EEG array. This limitation, however, is not as serious as it might appear, because most head translations are accompanied by rotations in practice. The rotation-based motion regressors might be able to efficiently reduce the motion-related variance in fMRI time courses even in the absence of the translational regressors. For example, the most significant motion, exhibited by Subject S1 (see Results), included both rotational and translational head movements. Yet, the E-REMCOR regressors, describing head rotations only, provided an efficient correction in this case (Fig. 4).

An important advantage of E-REMCOR is that it relies on the available and proven EEG-fMRI instrumentation and analysis techniques, and utilizes the rich head motion information already present in EEG-fMRI data. It can be applied retrospectively to any existing EEG-fMRI data set. Moreover, E-REMCOR should perform well if a simple head cap with several wire loops is used for motion detection instead of the EEG cap. This would simplify the experimental set-up, reduce preparation time, and allow application of E-REMCOR at those sites where the full-scale EEG-fMRI capability is not available.

An improved fMRI motion correction enabled by E-REMCOR should be particularly beneficial for fMRI at ultra-high magnetic fields, such as 7 tesla, because stronger magnetic susceptibility artifacts at higher fields make motion effects more pronounced. Also, many fMRI studies at 7 T use only limited spatial brain coverage, which may affect the quality of fMRI volume registration. Because motion artifact voltages in EEG recordings inside an MRI scanner are proportional to $B_0$, identification of motion-related ICs at 7 T will be easier and more accurate than at lower fields. E-REMCOR would also benefit the integration of fMRI with other neuroimaging modalities such as MEG (Zotev et al., 2008) and PET.

## 5. Conclusion

A novel method for retrospective motion correction of fMRI data using EEG is introduced. By utilizing motion artifacts in EEG recordings simultaneous with fMRI, E-REMCOR adds a high-resolution temporal dimension to the traditional fMRI motion correction, which relies on spatial registration of individual fMRI volumes. This additional temporal information makes it possible to reduce motion-related variance in fMRI data, particularly the effects of rapid head movements that cannot be adequately handled by the volume registration procedure. Thus, E-REMCOR bridges the gap between motion correction approaches in EEG and fMRI. It does not require any specialized equipment beyond the standard EEG-fMRI instrumentation and can be applied retrospectively to any existing EEG-fMRI data set. Because E-REMCOR regressors are based on EEG motion artifacts, they can be efficiently used to examine and reduce spurious motion-induced correlations between the EEG and fMRI data in simultaneous EEG-fMRI. From the clinical perspective, application of E-REMCOR can be expected to benefit fMRI data analysis for all subjects exhibiting significant head motions, including patients with neuropsychiatric disorders.

## Acknowledgments

This work was supported by the Laureate Institute for Brain Research and the William K. Warren Foundation. We would like to thank Dr. Robert Störmer, Dr. Patrick Britz, and Dr. Maria Schatt of Brain Products, GmbH for their continuous help and excellent technical support.